# *Single Photon Counting X-ray Imaging with Si and CdTe Single Chip Pixel Detectors and Multichip Pixel Modules*

Peter Fischer, Sven Krimmel, Hans Krüger, Markus Lindner,
Mario Löcker, Kazuhiro Nakazawa, Tadayuki Takahashi, Norbert Wermes

***Abstract*—Multichip modules (MCM) with 4 single photon counting MPEC 2.3 chips bump bonded to 1.3 cm x 1.3 cm large CdTe and Si semiconductor sensors as well as to single chip pixel detectors have been successfully built and operated.**

**The MPEC 2.3 chip provides a pixel count rate up to 1 MHz with a large dynamic range of 18 bit, 2 counters and energy windowing with continuously adjustable thresholds. Each MPEC has 32 x 32 pixels of 200 µm x 200 µm pixel size. For a MCM the 4 chips are arranged in a 2 x 2 array which leads to a 64 x 64 sensor pixel geometry. The MCM construction is described, and the imaging performance of the different detectors is shown. As readout system a newly developed USB system has been used.**

## I. Introduction

Single photon counting with hybrid pixel detectors has been introduced as a suitable method for direct digital imaging [1], [2]. Advantageous for a hybrid pixel detector is the variable use of different semiconductors as sensor material. As Si is a well understood semiconductor and homogeneous large Si wafers are easily available, Si is the preferred sensor material for applications where a high flux of ionizing radiation is used, e.g. material science or protein crystallography [3]-[8].

On the other hand, the low atomic number Z=14 of Si leads to a low absorption efficiency for X-rays with energies higher than a few 10 keV. This is not acceptable in low-dose medical imaging applications or hard X-ray and γ-ray astronomy. Thus, high Z-materials like GaAs (Z=31/33) were studied [9]-[11] and a very promising material for high absorption sensors is CdTe (Z=48/52). The absorption efficiency of an only 0.5mm thick CdTe sensor is 90% and 30% for X-ray energies of 40 keV and 100 keV, respectively. In addition, progress could be obtained in wafer production and electrode design so that high quality sensors with excellent spectroscopic performance are available [12]-[19]. As the connection of the readout chip to the fragile CdTe material is difficult, a sophisticated gold stud bump bond process was developed at ISAS (Institute of space and astronautical science, Japan) and a 0.5 mm thick CdTe sensor could be successfully bonded and operated with a previous version of the MPEC chip [20].

In order to use single photon counting detector systems for imaging applications, it is necessary to cover a large imaging plane which is achieved with multichip modules (MCM). This means that several readout chips are bump bonded to a single large sensor. In this work an arrangement of a square geometry with 2 x 2 MPEC chips was chosen and systems with a Si and a CdTe sensor each of those covering an active detection area of about 13 mm x 13 mm were successfully built and operated.

In this paper we present the MPEC 2.3 chip and report on studies with Si sensor MCMs and for the first time on measurements with single photon counting CdTe sensor MCMs.

## II. Pixel Detector description

### *A. Readout chip*

The MPEC 2.3 chip is a single photon counting pixel chip with 1 MHz high-count rate capability and energy windowing for photon energy discrimination. The active area of 6.4 mm x 6.4 mm is structured into 32 x 32 pixels of 200 µm x 200 µm size. Every pixel cell contains a preamplifier, two independent discriminators, and two 18 bit counters (Fig. 1). A coarse discriminator threshold is set globally, and a fine adjustment can be applied dynamically for each pixel. As both discriminators work independently from each other, an energy window can be set which can be useful in medical imaging for contrast enhancement [21]. The MPEC 2.3 chip is based on the MPEC 2.1 chip [22], and some crucial parts of the chip layout were reworked (counter, shift register, multiplexer, implementation of a new window-logic). The design of the MPEC 2.3 was done in the AMS 0.8µm technology.

H. Krüger, M. Löcker, N. Wermes are with Bonn University, 53115 Bonn, Germany (e-mail: loecker@physik.uni-bonn.de).

H. Krimmel was with Bonn University, 53115 Bonn, Germany. He is now with Technische Universität München, 80333 München, Germany.

M. Lindner was with Bonn University, 53115 Bonn, Germany. He is now with Robert Bosch GmbH, 71229 Leonberg, Germany.

K. Nakazawa, T. Takahashi, are with the Department of Physics, University of Tokyo, Bunkyo, Tokyo 113-0033 Japan, and also with the Institute of Space and Astronautical Science (ISAS), Sagamihara, Kanagawa 229-8510, Japan.

Unfortunately, this process offers only two metal layers which makes an efficient shielding between the digital part and the analog part of the chip difficult.

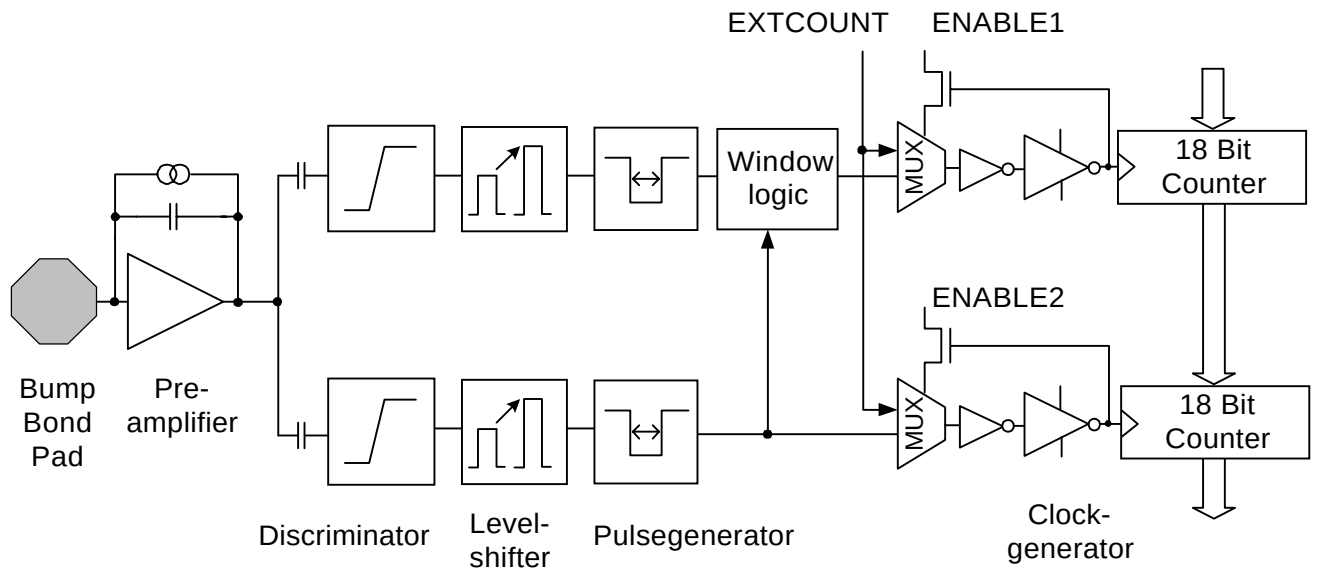

Fig. 1. Schematic of one pixel of the MPEC 2.3 chip.

## *B. The sensors*

For the presented measurements single chip sensors as well as MCM sensors were used. The 280 µm thick Si sensors are high resistance p+ on n sensors operated in diode configuration and the connection to the MPEC 2.3 chips is achieved by a wafer level solder bump bond process at IZM (Fraunhofer Institut für Zuverlässigkeit und Mikrointegration, Berlin, Germany) [23]. The p+ electrode is structured into the corresponding number of pixels of the readout chip, and all pixels are surrounded by a guard ring of 200 µm width.

The 0.5 mm thick high resistance CdTe sensors were fabricated by ACRORAD, Japan, and the detector assembling was done by ISAS. Both sides of the ohmic detectors are plated with Pt electrodes, and the side connected to the readout chip is structured into the corresponding number of pixels. Again, all pixels are surrounded by a guard ring of 200µm width. For the chip sensor connection a special gold stud bump bond technique in combination with an epoxy resin was used. More details of this sophisticated bonding procedure and characteristics of the used CdTe can be found in [20].

A critical issue for building MCM is the handling of the spacing between the readout chips (inter chip region). The readout chips have to be designed in a way that they are buttable at least at three sides, i.e. wire bond pads should be only on one chip side. Also, the sensor pixels which are connected to the corresponding pixels at the inner edges of the readout chips have to be larger than the normal pixel size because some additional space between the readout chips is required. Therefore, these pixels have a double side length in one direction leading to an area twice as large as the area of a normal pixel ("double pixel"). The four pixels in the center of the sensor have the double side length in both directions, thus, having a four times larger area ("quad pixel"). The arrangement of double and quad pixels can be seen schematically in Fig. 2. For the Si sensor a double pixel has a size of 200 µm x 400 µm (quad pixel: 400 µm x 400 µm) whereas for the CdTe sensor a double pixel geometry of 200 µm x 450 µm was chosen (quad pixel: 450 µm x 450 µm). Algebraically, the gap between the readout chips should be 400 µm (450 µm for the CdTe module), but as the readout chip has a surrounding margin, the effective gap between adjacent readout chips is only about 200 µm making MCM bump bonding challenging.

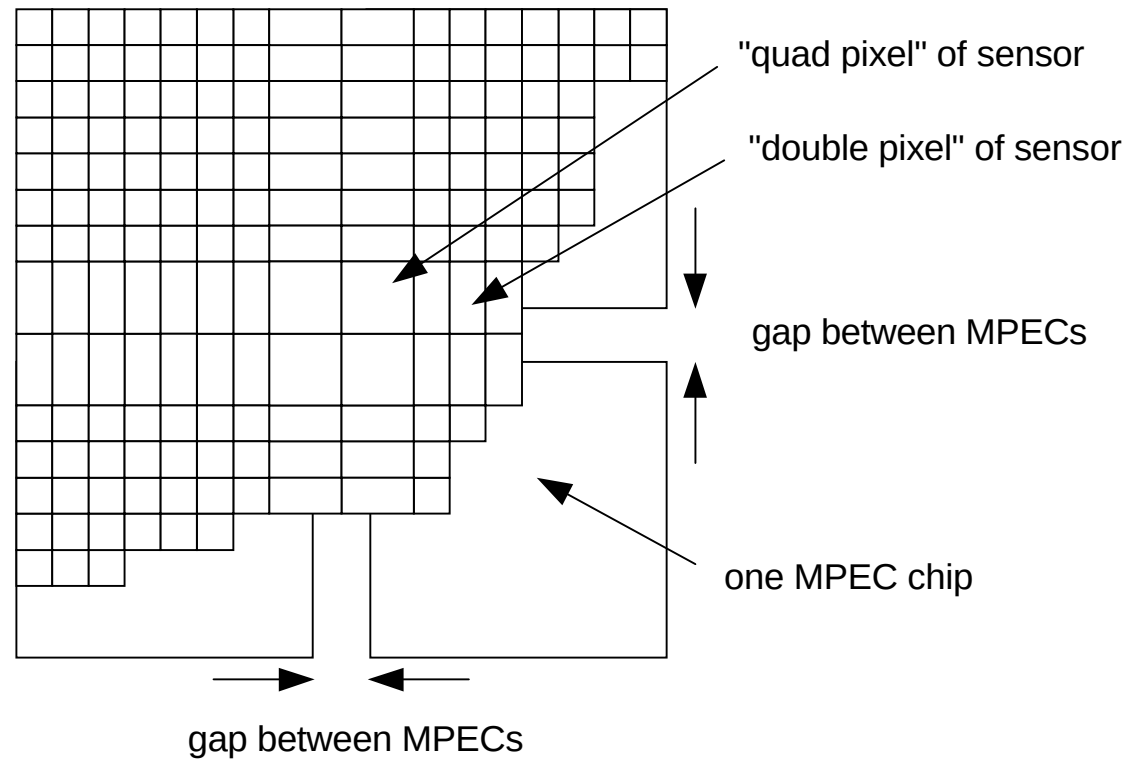

Fig. 2. Schematic view of a MCM with large pixels in the inter chip region.

## *C. Readout system*

All measurements with single chip detectors and with MCMs were done with a newly developed USB-based data acquisition system (s. Fig. 3).

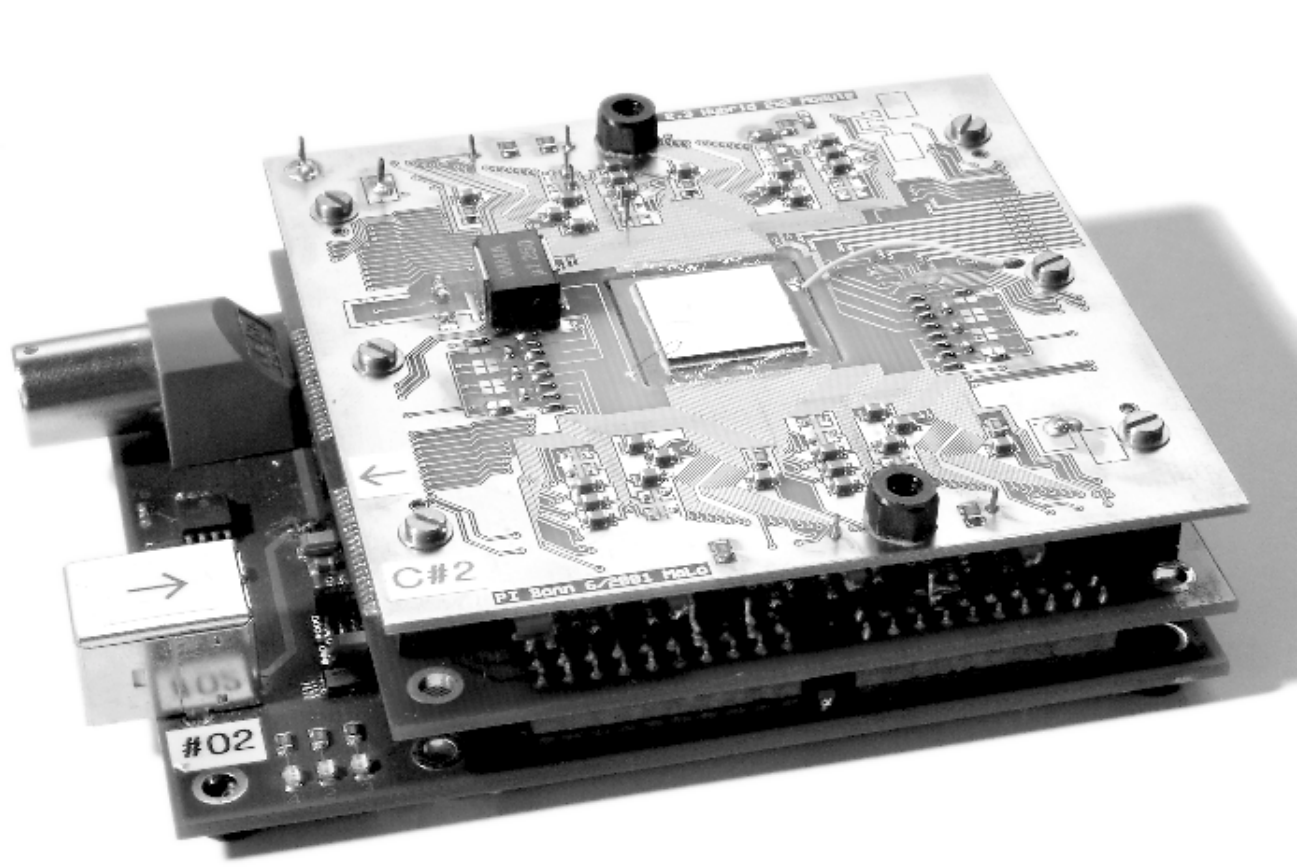

Fig. 3. Photograph of the USB-based detector system with a 2 x 2 CdTe sensor MCM (center of top PCB).

This small and handy system is controlled only by a laptop computer and is easily transported and set up. The detector can be self-powered by the USB-line, and with a battery stack for the detector depletion voltage the whole system is independent of any external power supply. Fig. 4 shows the compact stack arrangement of 3 PCBs. On the bottom the USB device controller card is placed performing the data exchange between the PC and the detector followed by the analog support card hosting all necessary analog and digital circuits for detector operation. On top, the hybrid adapter card is mounted by a zero force connector.

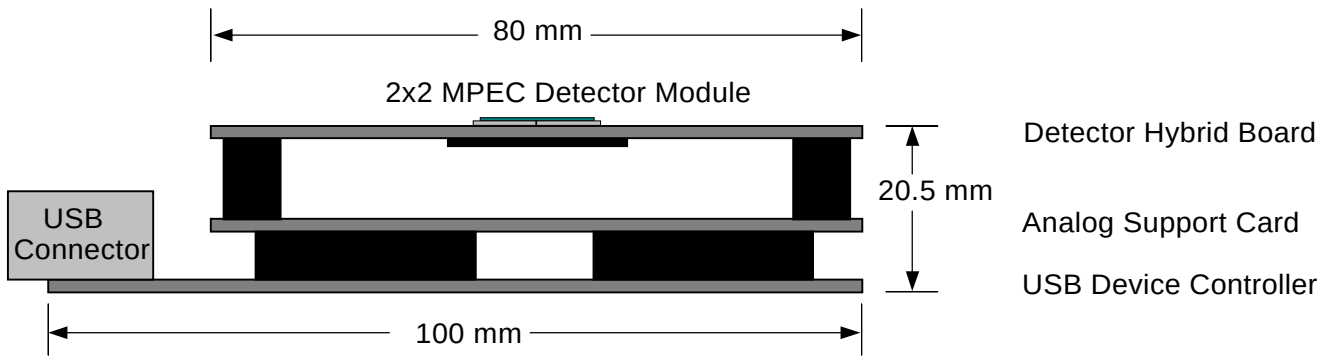

Fig. 4. Schematic view of the USB readout system for MCMs and single chip detectors. A stack of three PCBs makes the system compact and mobile.

In order to avoid mechanical stress due to shear forces originating from the mismatch of thermal expansion coefficients or external forces, the detector module itself is not fixed to the hybrid adapter card but to a small metalized ceramic plate. This ceramic plate in turn is put through a corresponding opening in the hybrid adapter card and fixed with another ceramic plate (s. Fig. 5). The use of ceramic material ensures a good match of the thermal expansion coefficient of Si and a good heat transfer during chip operation.

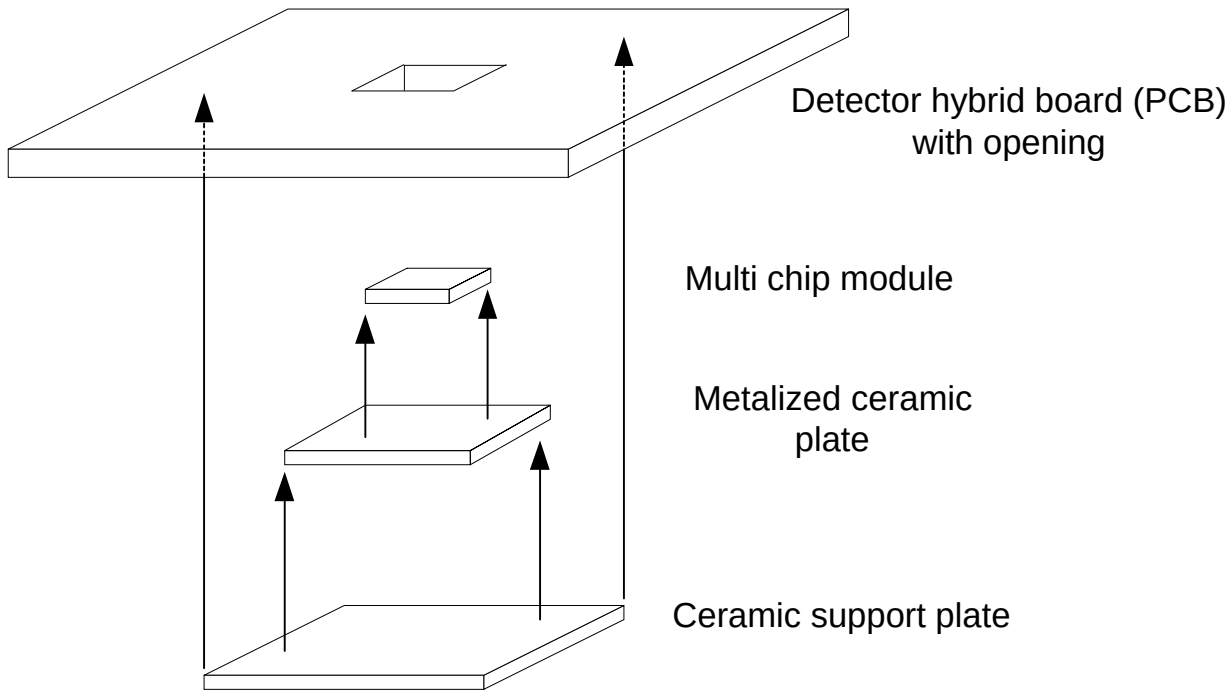

Fig. 5. Arrangement of the ceramic plates connecting the MCM to the detector hybrid board.

## III. MEASUREMENTS WITH SINGLE CHIP DETECTORS

### A. Thresholds

The MPEC 2.3 is a low noise ASIC whose equivalent noise charge (ENC) is measured to be about 60 e without sensor and about 110 e with additional input capacitance of the sensor. The minimum threshold can be set to about 1300 e. This high threshold has to be set to avoid digital analog cross talk and is not in contradiction to the low noise performance of the chip. The digital switching noise couples via the chip bulk and the sensor into the preamplifier and would cause false signals if the threshold is not high enough. As mentioned earlier, an effective shielding could not be achieved in the AMS 0.8µm technology where only two metal layers are available.

An important characteristic for the imaging performance of a photon counting chip is a low dispersion of the thresholds of all pixels in order to precisely setting energy cuts. In the case of the MPEC 2.3 chip both thresholds of a single pixel can be independently adjusted by storing a correction voltage on a capacitor in each pixel. As a result, the RMS-value of the threshold dispersion without sensor could be reduced from 180 e for the unadjusted case to 10 e for the adjusted case (Fig. 6 and 7). A low threshold dispersion could already be demonstrated with the MPEC 2.1 in [22]. However, the outer columns of the MPEC 2.1 chip did not work properly whereas this problem is eliminated for the MPEC 2.3.

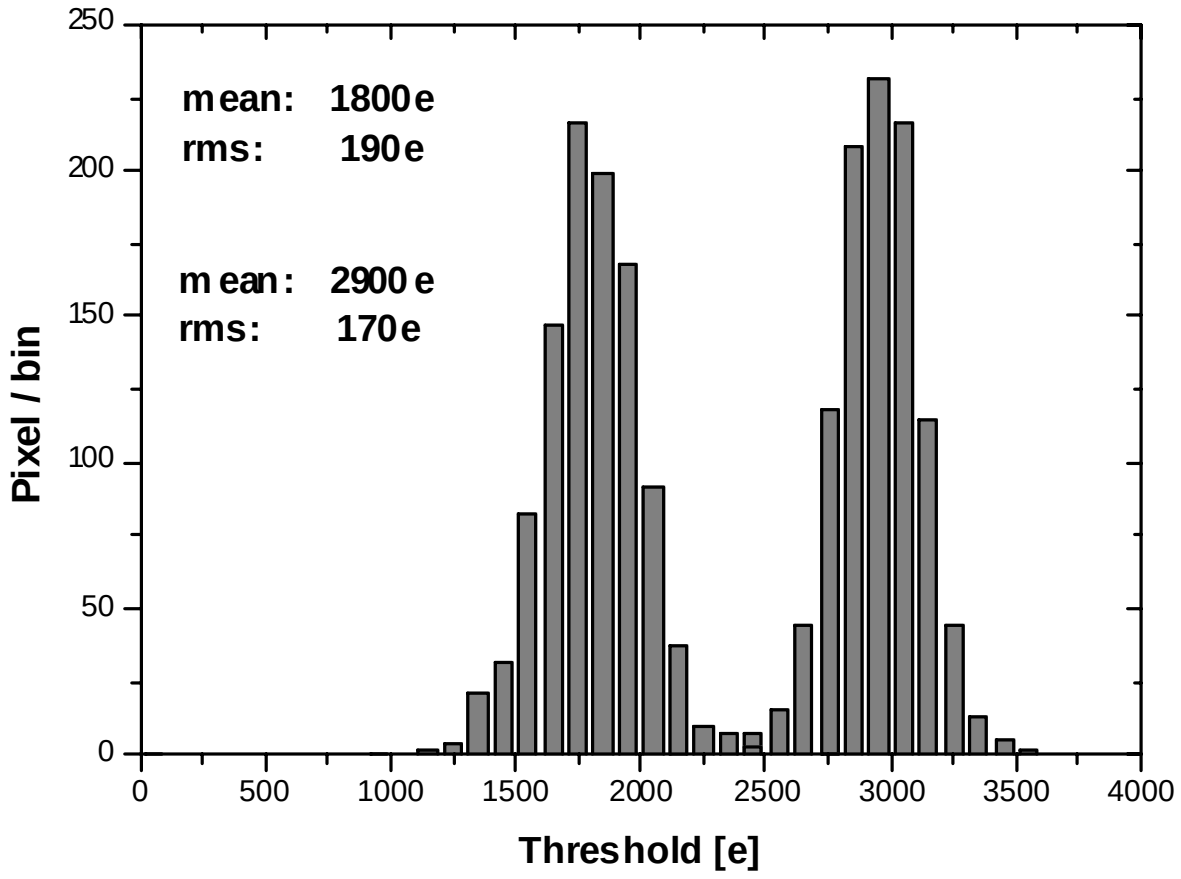

Fig. 6. Dispersion of unadjusted lower and upper thresholds of MPEC2.3 (without sensor).

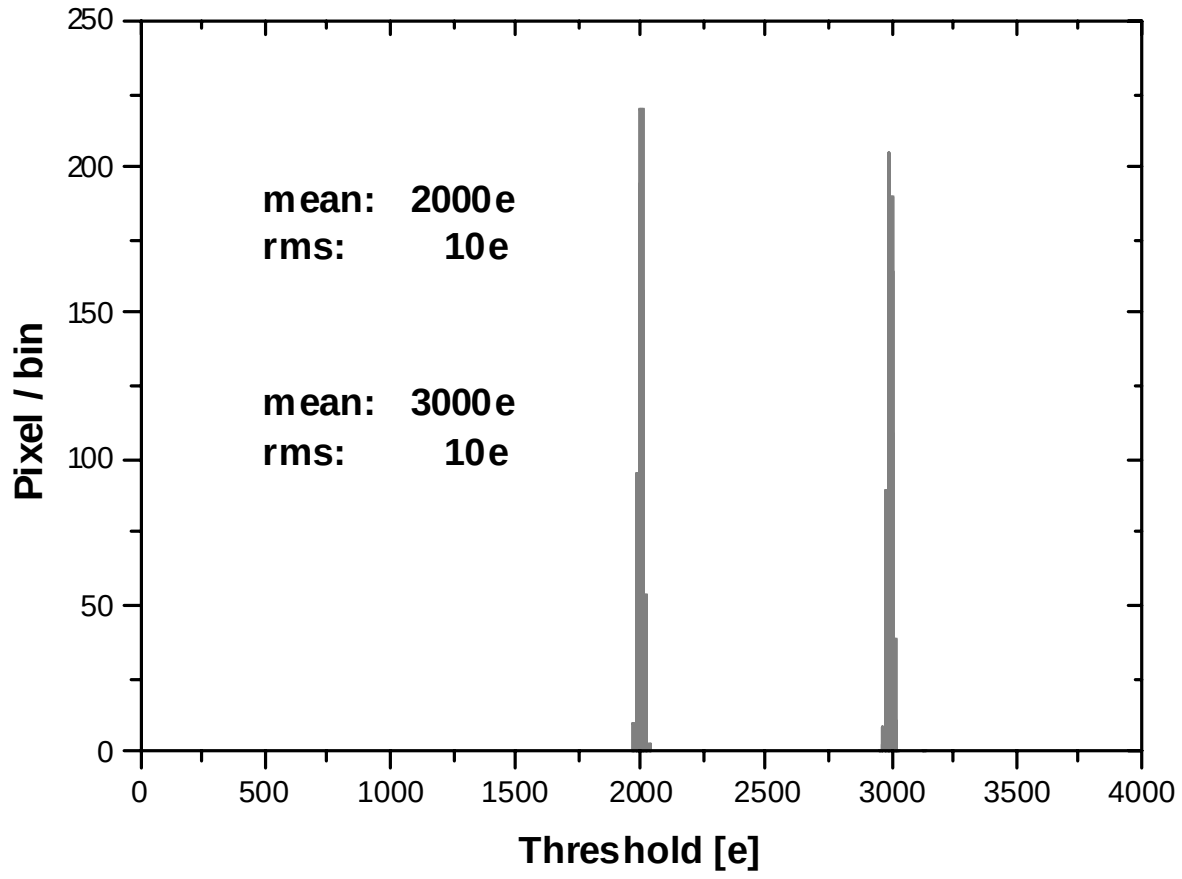

Fig. 7. Dispersion of adjusted lower and upper thresholds of MPEC2.3 (without sensor).

As the threshold adjustment is done dynamically, the correction voltage needs to be refreshed. In order to minimize the refresh rate, a threshold drift compensation circuit was developed reducing the drift current from the capacitor [22]. As an example, the measured drift of the lower threshold of the 32 pixels of one column is shown in Fig. 8. At the beginning of the measurement the thresholds were set to 3000 ± 10 e and then were measured every 2 minutes. It turns out that the maximum drift rate is < 0.2 e/s so that only one fine adjustment cycle is needed for a typical detector exposure time.

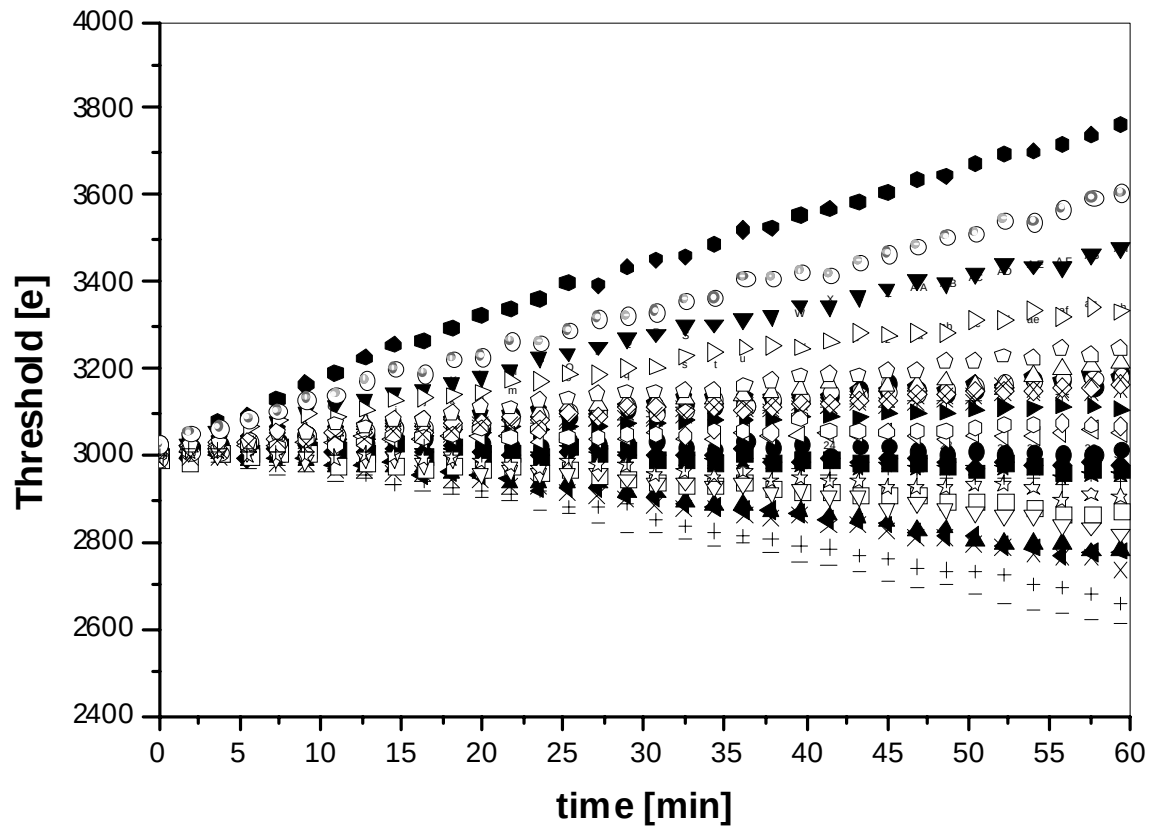

Fig. 8. Threshold drift of 32 pixels within one hour.

## B. *Imaging*

The digital analog cross talk limiting the threshold settings is not only caused by the counter flip-flops of a pixel cell but also by the switching noise of digital control lines. The influence of the digital control lines could be significantly reduced by a modified readout sequence of the chip so that the chip could be operated with a threshold of about 1300 e. This is demonstrated by a flat field image of a Si sensor taken with 6 keV photons of $^{55}$Fe shown by Fig. 9 (6 keV energy deposition corresponds to approx. 1600 electron-hole pairs in Si).

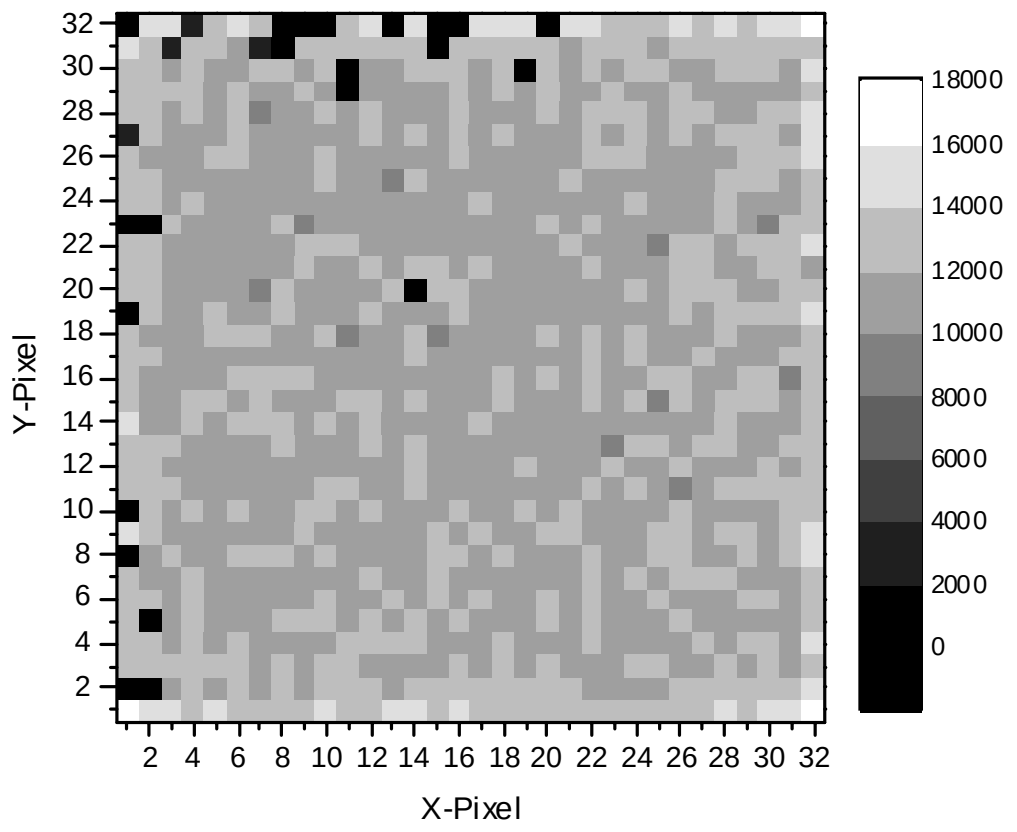

Fig. 9. Flat field image of a Si sensor taken with 6 keV photons of $^{55}$Fe (adjusted threshold set to 1300e).

Also, a relatively good bump yield can be seen from Fig. 9. Apart the periphery of the chip only a few pixels are not successfully connected to the sensor. An example for a radiograph of a low absorbing object shows Fig. 10, where a mosquito is exposed to 6 keV photons of $^{55}$Fe.

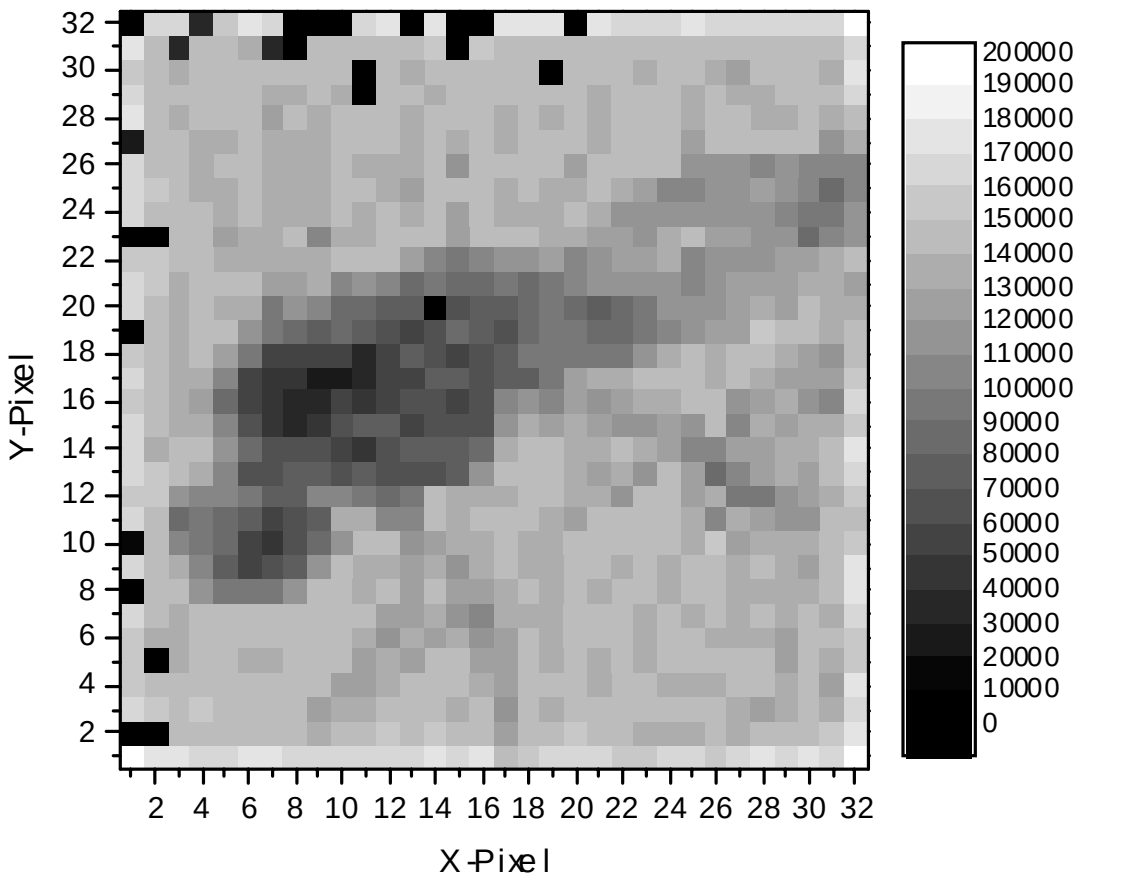

Fig. 10. Radiograph of a mosquito taken with 6 keV photons of $^{55}$Fe (Si sensor, adjusted threshold set to 1300e).

A flat field image taken with a CdTe sensor is depicted in Fig. 11. Only 4 pixels out of 1024 are not connected to the sensor which proves an excellent bump yield achieved with gold stud bump bonding. However, in the flat field image a column pattern of the count rate distribution becomes visible. As the CdTe sensor is symmetric row- and column-wise, this phenomenon must be induced by the readout chip in combination with the gold studs arrangement which indeed has a column dependence. The exact origin of this non-uniform behavior is still under investigation. Nevertheless, the detector is fully functional which is demonstrated in Fig. 12 by a radiograph of a screw nut taken with 60 keV photons of $^{241}$Am.

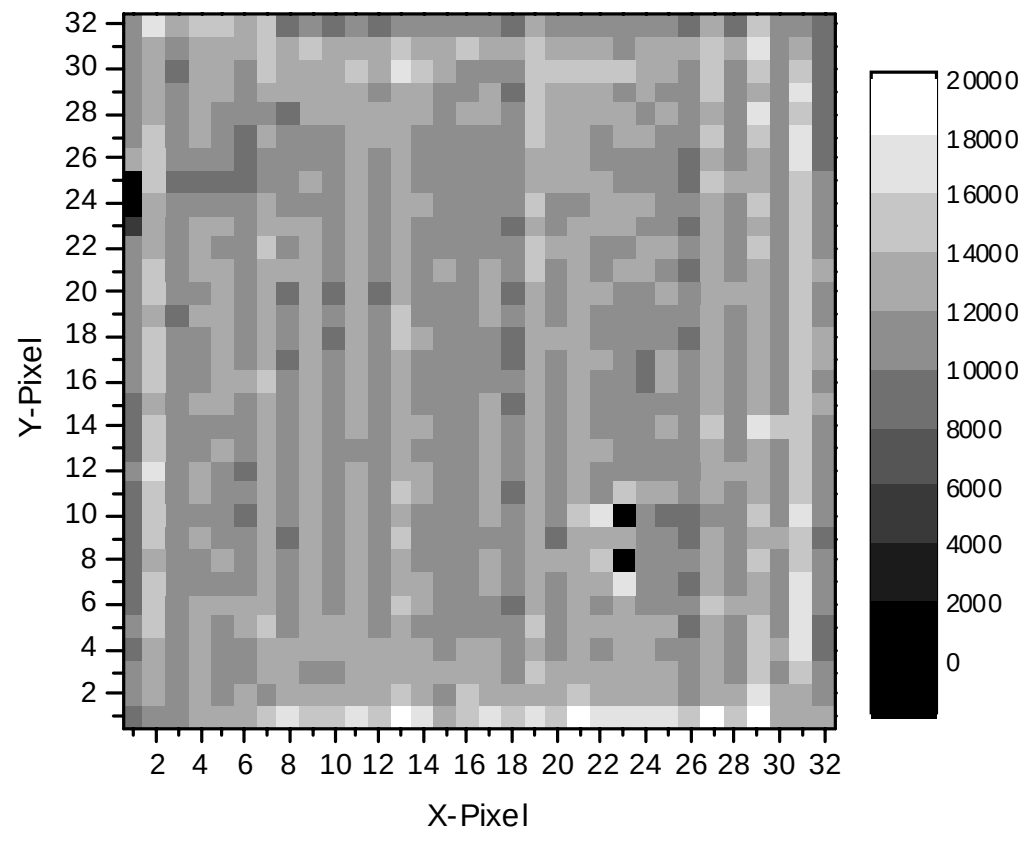

Fig. 11. Flat field image of a CdTe sensor taken with 60 keV photons of $^{241}$Am (without threshold adjustment).

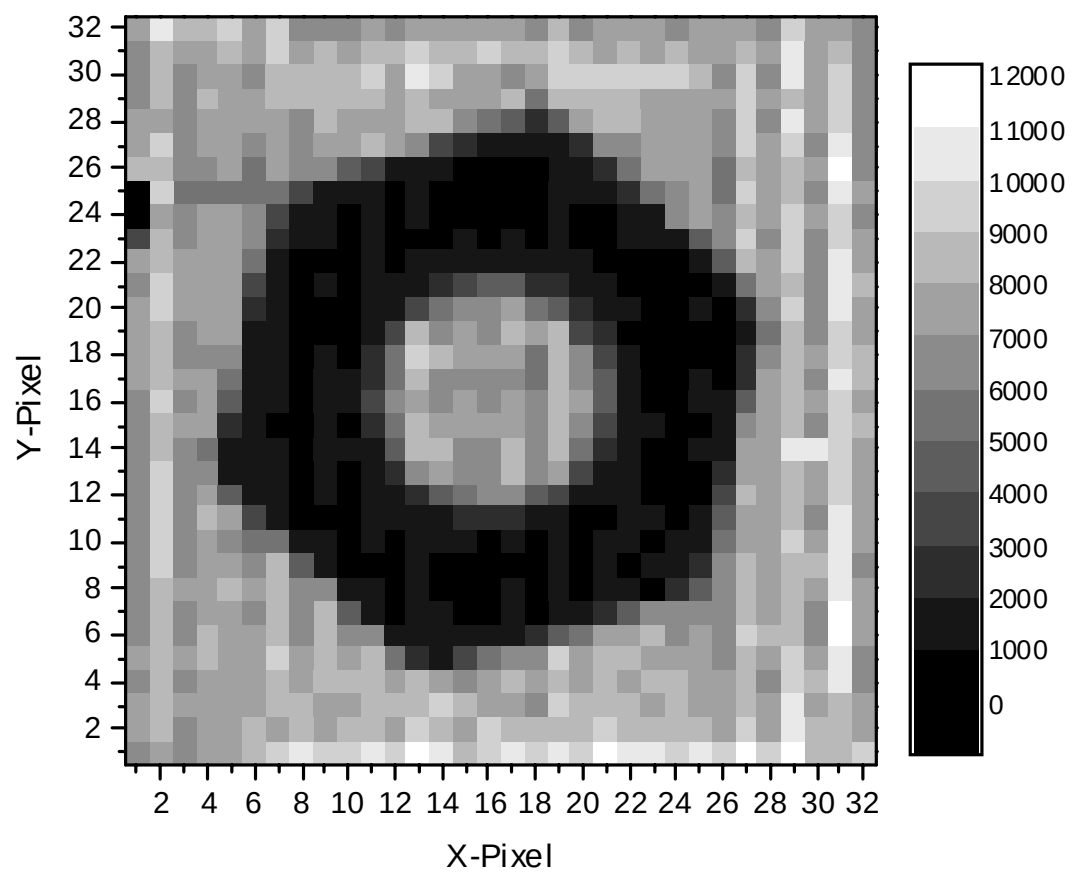

Fig. 12. Radiograph of a screw nut taken with 60 keV photons of $^{241}$Am and CdTe sensor (without threshold adjustment).

## IV. Measurements with Multi Chip Modules

### A. Si sensor MCM and CdTe sensor MCM

Available bump bonding techniques for MCM construction are wafer level processes based on In or PbSn bumps, and with these methods Si sensor MCMs for high energy physics experiments have been successfully built [23-26]. For the Si sensor MCM described in this work a PbSn bump bonding process was used. As the MPEC 2.3 chips were only available as single dies, the MPEC chips had to be inserted into a dummy wafer, in which an exactly matched opening was cut before. After wafer processing the chips were removed out of the dummy wafer again and cleaned along their edges. Although this whole procedure degraded the bump yield along the chip edges, the built Si sensor MCM could be successfully operated.

For future biomedical applications the challenge is to build MCMs with large area CdTe sensors. In contrast to the above mentioned techniques the CdTe sensor MCM is bump bonded by a gold stud process. This process does not require a full wafer but can be done with single chip dies. A flat field image taken with a CdTe MCM is shown by Fig. 14.

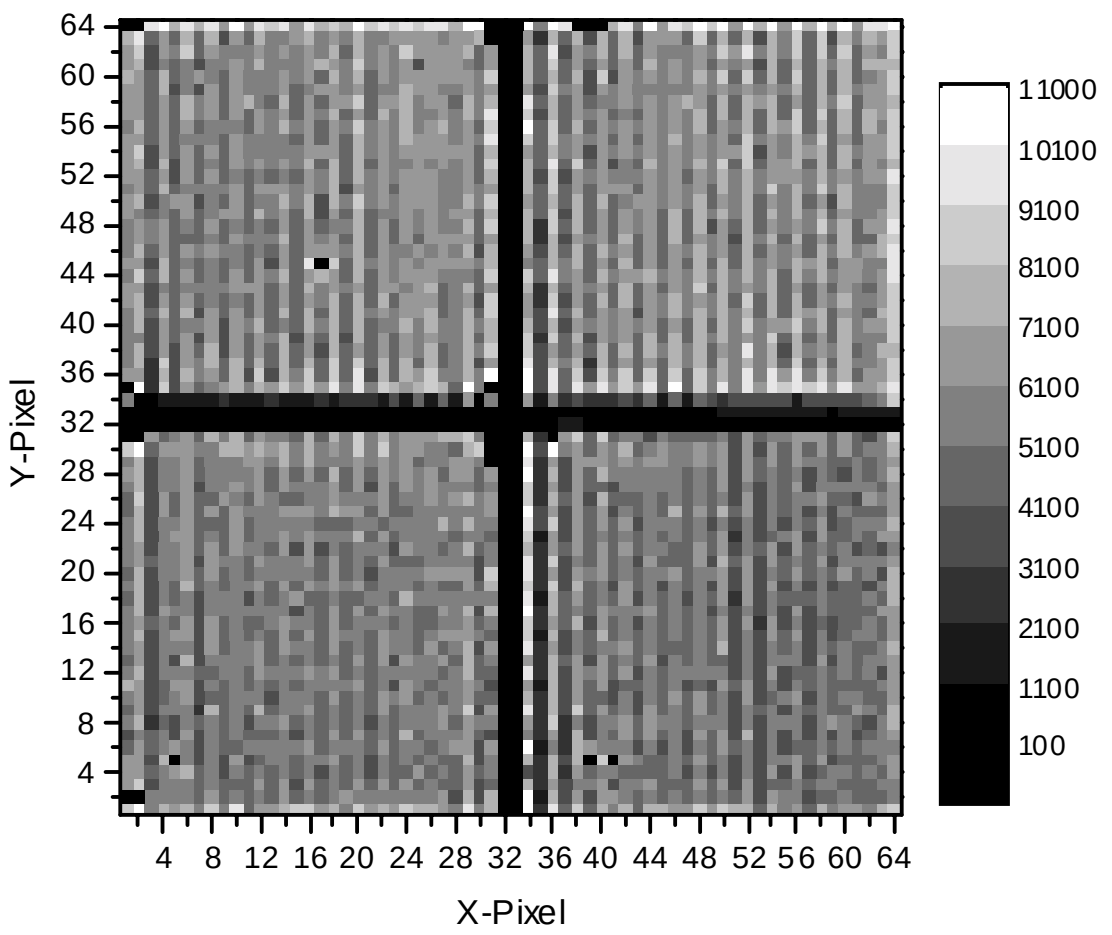

Fig. 14. Flat field image of a CdTe sensor MCM (without threshold adjustment).

It can be seen that along the two axes of the module many defect pixels exist. These defect pixels have either a completely vanishing or a very low count rate. Here, a possible explanation is a break of the detector along the two axes damaging the electrodes of the adjacent pixels. If a part of the broken electrode is still connected to the readout pixel, the measured count rate is reduced according to the reduced electrode size. A break of the CdTe is not unlikely because the very fragile 0.5 mm thick sensor is mechanically stressed during bump bonding to the readout chips and during wire bonding to the hybrid PCB. But it should be stated that all other regions of the MCM show a very good bump yield. On the other hand, the column pattern in the count rate distribution can be recognized again which was already observed with the CdTe single chip detector described above. An example of the imaging performance of the CdTe sensor MCM is presented in Fig. 15 showing the radiograph of a cogwheel.

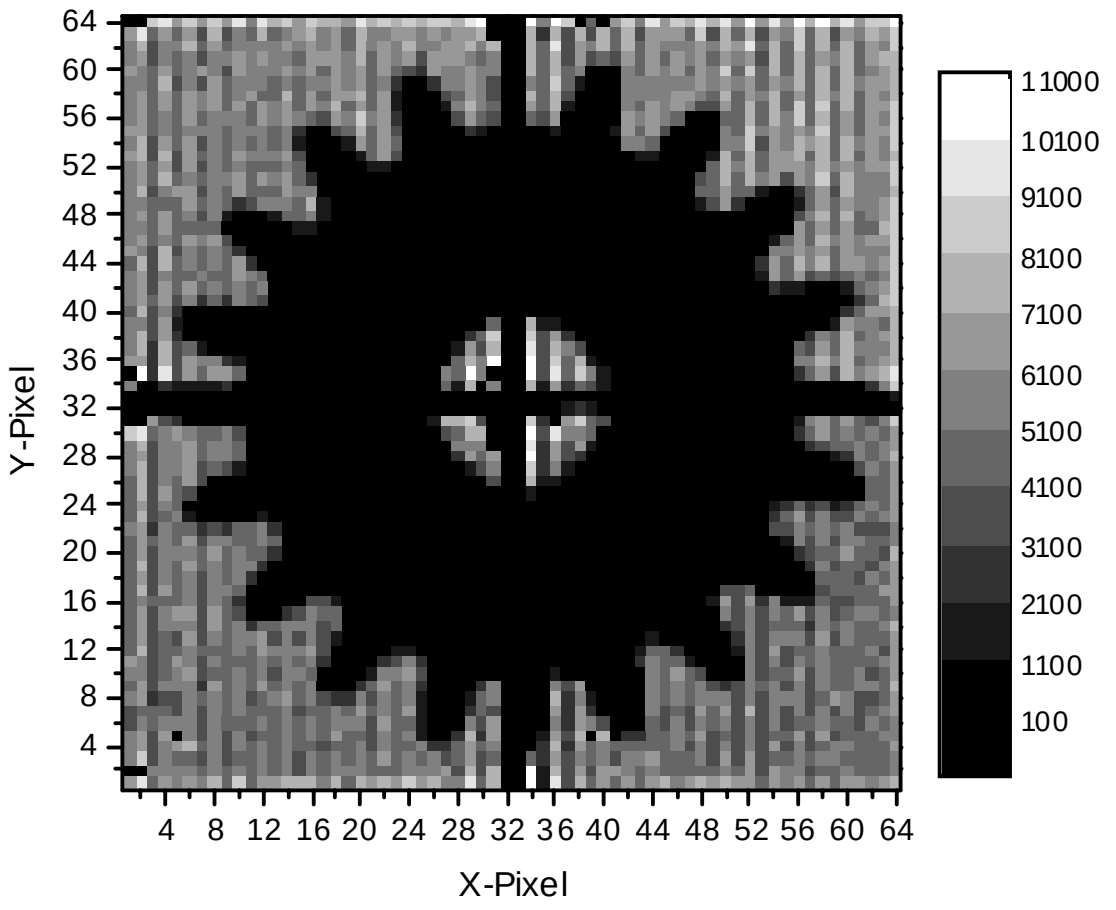

Fig. 15. Radiograph of a cogwheel taken with 60 keV photons of $^{241}$Am and CdTe sensor MCM (without threshold adjustment).

## V. SUMMARY

The MPEC 2.3 chip could successfully characterized as low noise single photon counting read out chip for pixel detectors. By dynamically adjusting the thresholds a very low threshold dispersion could be achieved. A single chip Si detector of 300 µm thickness and a single CdTe detector of 500 µm thickness were assembled and operated. In addition a Si sensor MCM and for the first time a photon counting MCM with a CdTe sensor were built. Both, solder bump bonding for the Si sensors and gold stud bump bonding for the CdTe sensors were utilized.

The solder bump bonding is a wafer level based process for which the single diced MPEC chips have to be inserted into dummy wafers. The gold stud bump bonding shows especially in the case of a single chip detector a very good bump yield. But it also should be noted that the imaging performance is somehow affected by a column dependent count rate distribution. As for the assembling of the CdTe sensor MCM special care must be taken when applying mechanical forces to avoid a break of the thin sensor.

All measurements were performed with a newly developed USB based readout system. Its compact size and self-powering possibility makes the whole system easy to use and versatile for different applications.

## VI. ACKNOWLEDGMENT

We would like to thank W. Ockenfels and Ö. Runolfsson for their help in detector assembling.